\documentclass[]{spie}  

\usepackage{amsmath,amsfonts,amssymb}
\usepackage{graphicx}
\usepackage[colorlinks=true, allcolors=blue]{hyperref}
\usepackage{subfig}

\title{Building a GRAVITY+ Adaptive Optics Test Bench}

\author[a]{The GRAVITY+ consortium: Florentin Millour}
\author[a]{Philippe Berio}
\author[a]{St\'ephane Lagarde}
\author[a]{Sylvie Robbe-Dubois}
\author[a]{Carole Gouvret}
\author[a]{Olivier Lai}
\author[a]{Fatmé Allouche}
\author[a]{Christophe Bailet}
\author[a]{Olivier Boebion}
\author[a]{Marcel Carbillet}
\author[a]{Aurélie Marcotto}
\author[a]{Alain Spang}
\author[b]{Paul Girard}
\author[b]{Nicolas Mauclert}
\author[c]{Jean-Baptiste Lebouquin}
\author[d]{Thibaut Paumard}
\author[e]{Ferreol Soulez}
\author[f]{Julien Woillez}
\author[g]{Nikhil More}
\author[g]{Frank Eisenhauer}
\author[h]{Christian Straubmeier}
\author[i]{Laura Kreidberg}
\author[j]{Paulo Garcia}
\author[k]{Sebastian Hoenig}

\affil[a]{{\small Universit\'e C\^ote d'Azur, Observatoire de la C\^ote d'Azur, CNRS, Laboratoire Lagrange, France}}
\affil[b]{{\small Universit\'e C\^ote d'Azur, Observatoire de la C\^ote d'Azur, CNRS, UMS Galilée, France}}
\affil[c]{{\small Univ. Grenoble Alpes, CNRS, IPAG, 38000 Grenoble, France}}
\affil[d]{{\small LESIA, Observatoire de Paris, Universit\'e PSL, CNRS, Sorbonne Universit\'e, Universit\'e de Paris, 5 place Jules Janssen, 92195
Meudon, France}}
\affil[e]{{\small Univ Lyon, Univ Lyon1, Ens de Lyon, CNRS, Centre de Recherche Astrophysique de Lyon UMR5574, F-69230, Saint-Genis-Laval, France}}
\affil[f]{{\small European Southern Observatory, Karl-Schwarzschild-Straße 2, 85748 Garching, Germany}}
\affil[g]{{\small Max Planck Institute for extraterrestrial Physics, Giessenbach-straße 1, 85748 Garching, Germany}}
\affil[h]{{\small 1st Institute of Physics, University of Cologne, Z\"ulpicher Straße 77, 50937 Cologne, Germany}}
\affil[i]{{\small Max Planck Institute for Astronomy, Königstuhl 17, 69117 Heidelberg, Germany}}
\affil[j]{{\small Faculdade de Engenharia, Universidade do Porto, rua Dr. Roberto Frias, 4200-465 Porto, Portugal}}
\affil[k]{{\small Department of Physics \& Astronomy, University of Southampton, Southampton, SO17 1BJ, UK}}

\authorinfo{Further author information: \\ Florentin Millour: E-mail: florentin.millour@oca.eu, Telephone: +33 (0)6 98 02 44 51}

\begin{document} 
\maketitle

\begin{abstract}
We present the testbench aimed at integrating the GRAVITY+ adaptive optics GPAO. It consists of two independent elements, one reproducing the Coudé focus of the telescope, including the telescope deformable mirror mount (with its surface facing down), and one reproducing the Coudé room opto-mechanical environment, including a downwards-propagating beam, and the telescope mechanical interfaces in order to fit in the new GPAO wavefront sensor. We discuss in this paper the design of this bench and the solutions we adopted to keep the cost low, keep the design compact (allowing it to be fully contained in a 20 sqm clean room), and align the bench independently from the adaptive optics. We also discuss the features we have set in this bench.
    
\end{abstract}


\section{INTRODUCTION} \label{sec:intro} 

GRAVITY+ \cite{GRAVITY+22} is an update of the GRAVITY instrument in particular and of the facilities of the Very Large Telescope Interferometer in general (an observatory of ESO in Chile). Breakthrough results have been obtained with GRAVITY on the Galactic Center\cite{GRAVITY22}, active galactic nuclei\cite{2018Natur.563..657G}, and exoplanets\cite{2020A&A...633A.110G}. The objective of GRAVITY+ is to go beyond these  results by enabling wide-field off-axis fringe tracking ($\approx 30''$). This will improve the sensitivity and the contrast of GRAVITY, enabling the detection of objects orbiting 10x closer SgrA* (mag$K \leq 22$ vs $19$ for broadband imaging, and $V \geq 0.75$ vs $0.45$, respectively). This goal will be achieved with new natural guide star and laser guide star adaptive optics (NGS-AO and LGS-AO, respectively) installed at the four VLT 8-m Unit Telescopes (UT).

The GRAVITY+ Adaptive Optics (GPAO) is a major work package of GRAVITY+. The work is structured around a system team composed of INSU (IPAG, Grenoble, LESIA, Paris and Lagrange, Nice), and MPE, Garching, and each institute has the responsibility of an AO subsystem and of the test plan: IPAG, Grenoble and MPE, Garching have the responsibility of the ALPAO 41x41 Deformable Mirror (DM), LESIA, Paris takes care of the Real Time Computer (RTC), MPE is in charge of the Wavefront Sensor (WFS) and the control and command software (ICS), and finally Lagrange, Nice leads the AO test plan (AIT) with an AO test bench which mimics the optical interfaces of the VLT, and which will be used to perform the AO functional testings.
The first AO system will be installed in Nice in 2022. It is presently planned to install the first system at Paranal in 2024, and to use laser guide stars in 2025.

\section{GPAO testbench presentation} \label{sec:presentation} 

\begin{figure}[htbp]
    \centering
    \includegraphics[width=0.45\textwidth]{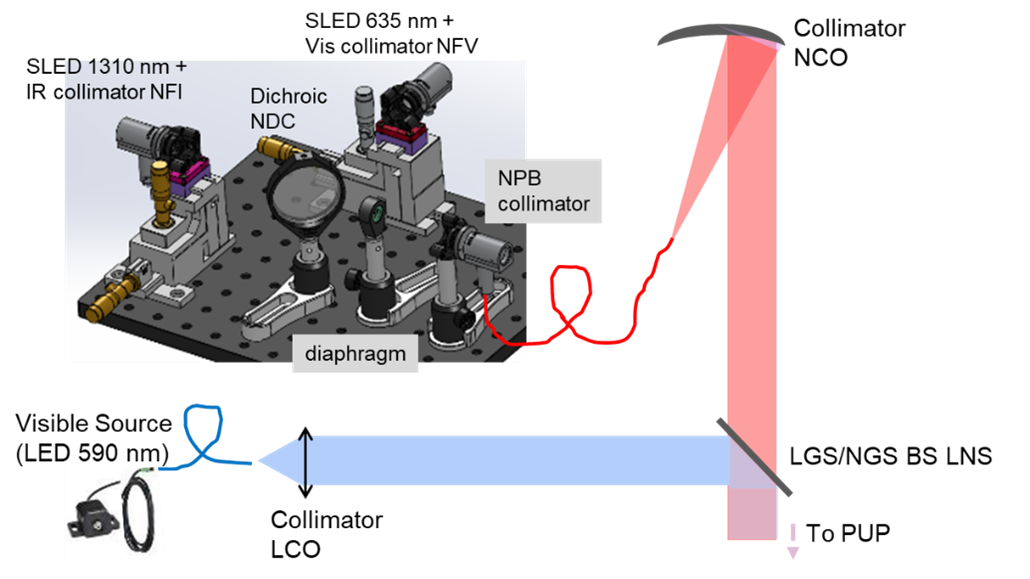}
    \includegraphics[width=0.45\textwidth]{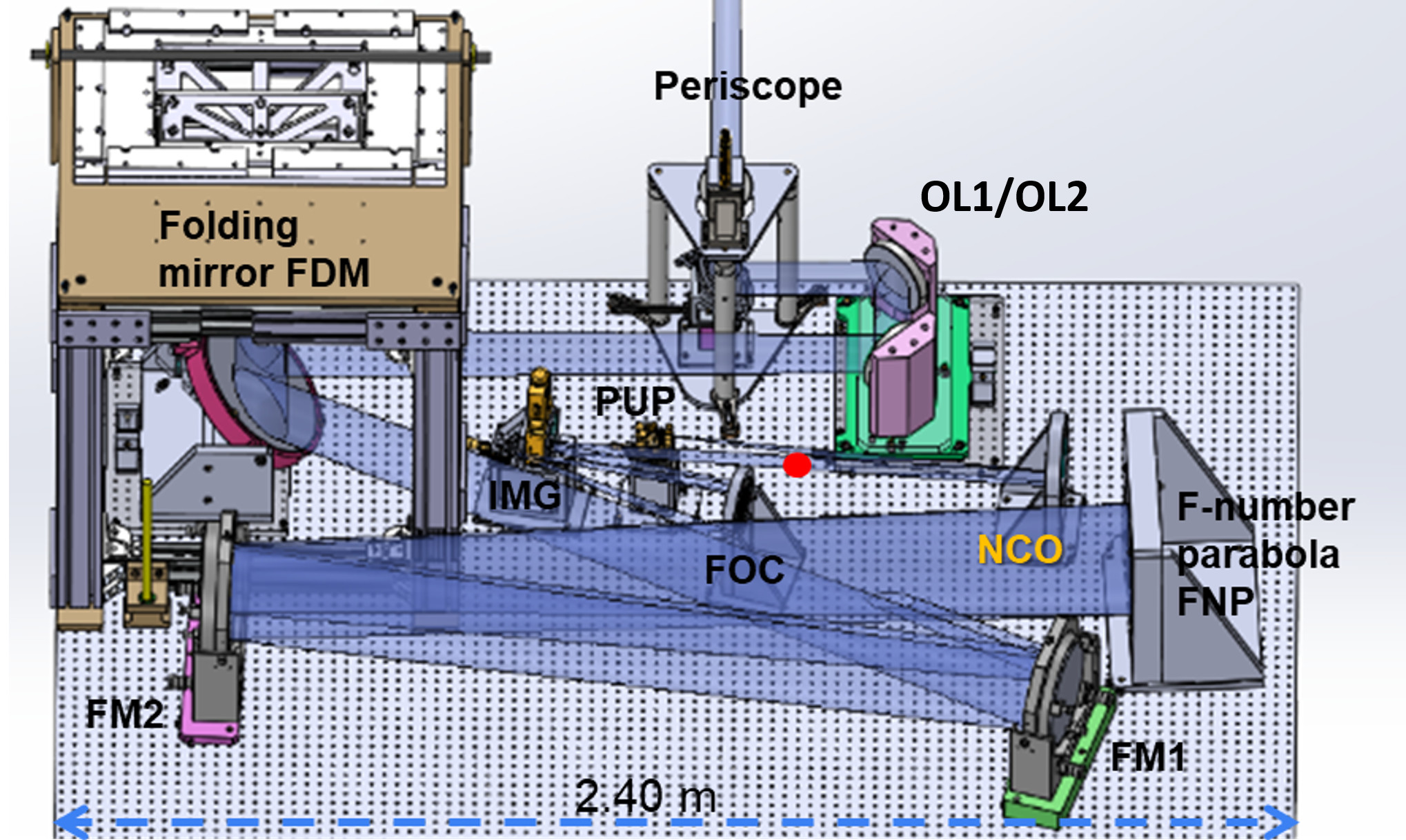}
    \includegraphics[width=0.45\textwidth]{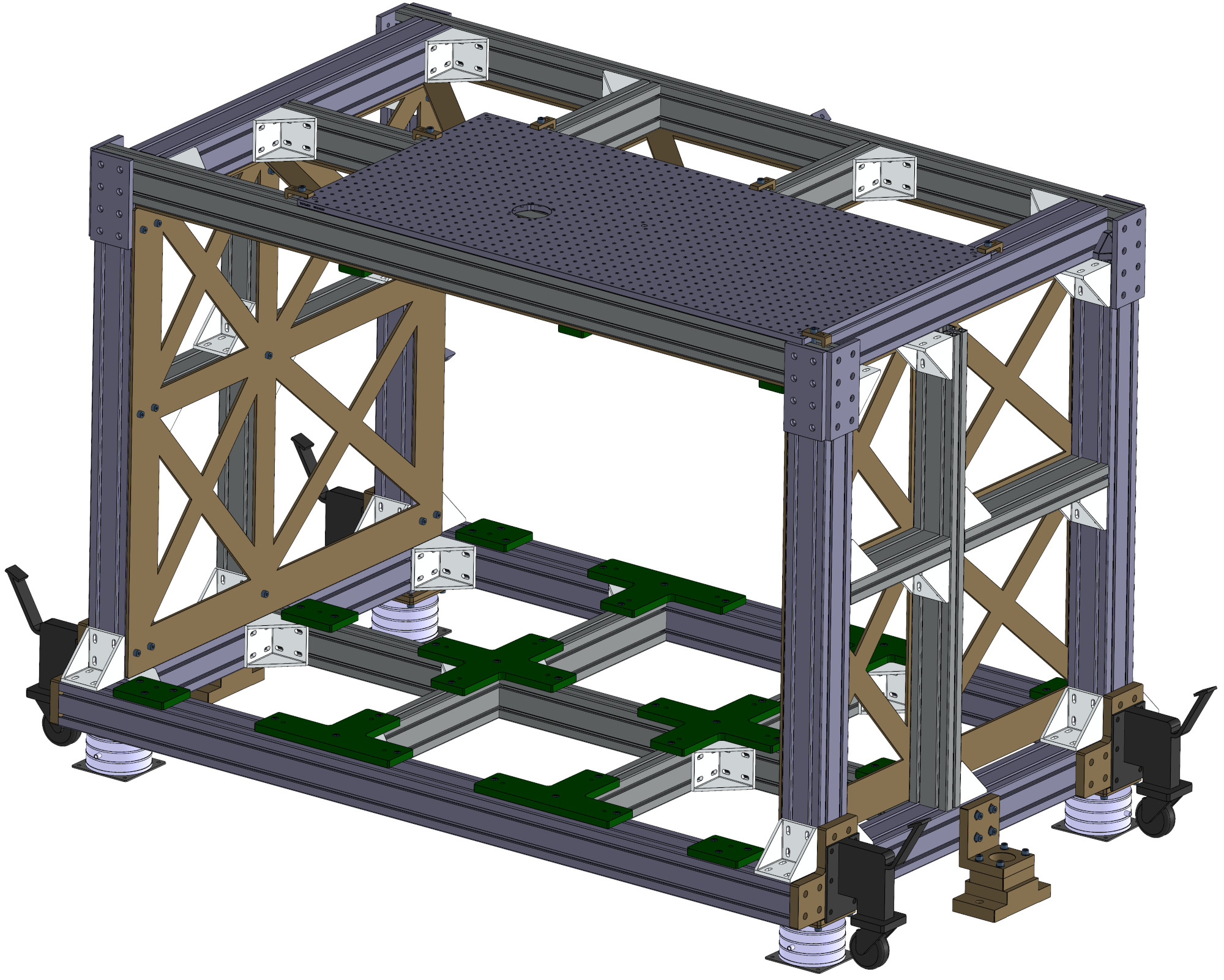}
    \includegraphics[width=0.45\textwidth]{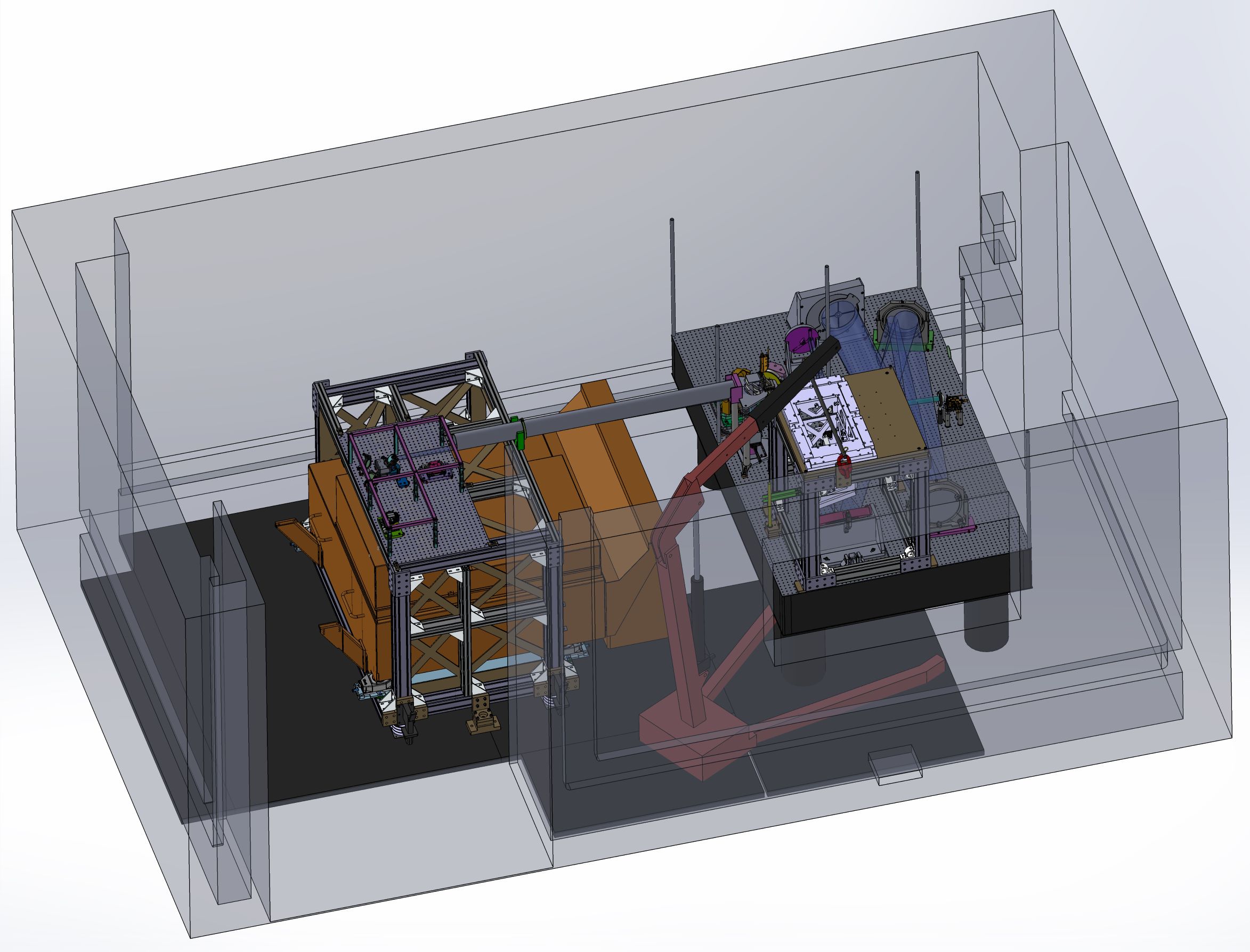}
    \caption{{\bf Top-Left:} CAD model of the testbench NGS light source and a sketch of the LGS light source. {\bf Top-Right:} GPAO test bench telescope simulator CAD view, with the Corrective Optics support structure on the top-left. Optical elements have all a 3-letter acronym recalling their functions (see text for details). {\bf Bottom-Left:} WFS support stucture CAD view.  {\bf Bottom-Right:} The GPAO testbench in its clean room.}
    \label{fig:schematics}
\end{figure}

The GPAO testbench is mostly inspired from the MACAO testbench\cite{MACAO_AIT}, with a source, a telescope simulator (which includes a Corrective Optics support structure), and a Wavefront Sensor support structure. 

The source is designed to provide a "Natural Guide Star" (NGS), i.e. a photonic crystal monomode fiber fed by two superluminous LEDs (one visible and one infrared) and a "Laser Guide Star" (LGS) that is injected directly from a multimode fiber (to simulate an extended source) into the collimated beam of the telescope simulator (see Figure \ref{fig:schematics}, top-left).

The telescope simulator (Figure \ref{fig:schematics}, top-right) provides the same incoming light beam as in a Unit Telescope, i.e. a vertical beam of 10cm diameter onto the deformable mirror, inclined at 10 degrees, and a F-number of 46.5. It also provides degrees of freedom to shift the pupil laterally, to move the source in the field of view, and allows us to place a rotating phase plate in reflection to simulate atmospheric turbulence.

The support structure of the Corrective Optics (Figure \ref{fig:schematics}, top-left corner in the top-right picture) and Wavefront Sensor support structure (Figure \ref{fig:schematics}, bottom-left) provide the exact same mechanical interfaces as the inner azimuth track and the Coudé focus of the Unit Telescope, respectively. The large WFS structure is installed aside the optical table, while the Corrective Optics support structure could be fit onto the main optical table.

We make use of one optical element of the MACAO testbench: its telecentricity lens, located just above the WFS to provide a flat field of view to the WFS. The bench is installed on a 1m50 x 2m40 optical table in a clean room of the Lagrange laboratory (Fizeau building, figure \ref{fig:schematics} bottom-right), and we took profit of the copious amount of flat mirrors needed to fold the beam onto such a small optical table (7 flats) to provide additional adjustments between the telescope simulator and the WFS support structure (namely: tilt and lateral adjustment of the beam with a periscope, and beam focus with two translating folding mirrors OL1/OL2).

The beam propagates starting from the source fiber output (SFO, red point in the top-right panel). The light first hits the NCO (collimator) parabolic mirror that prouces a parallel beam. It then goes to the PUP (pupil) flat mirror. PUP is where the pupil mask reproducing the pupil of the VLT with spiders and central obstruction is placed. It is in a conjugated plane with the DM (Deformable Mirror). Then the beam hits the FOC focusing parabolic mirror before the IMG flat mirror, where an image of the fiber is formed. After IMG, which adjusts by tilt the beam lateral position (pupil shift), the beam continues through FM1 and FM2 (folding flat mirrors) up to the large FNP parabolic mirror (F-Number Parabola). Adjusting the distance between IMG and FNP allows the F-number adjustment.

After the FNP parabola, the beam hits back FM2, then FM1, and encounters the flat FDM folding mirror, that reflects the inclined vertical beam towards the DM. The 10\,cm converging beam is then reflected vertically by the DM, hits FDM a second time, and propagates horizontally to the OL1/OL2 (focus/optical length adjustment device) and then the periscope (providing an interface between the telescope simulator and the separated WFS structure).

The testbench also provides two imaging focuses (IR and visible) located onto a breadboard installed on top of the WFS structure after picking part of the incoming flux with a 50/50 beamsplitter. It allows us to visualize either the image or the pupil of the simulator, in a very similar way as a science camera mounted on an adaptive optics.

\section{GPAO testbench status} \label{sec:status}

\begin{figure}[htbp]
    \centering
    \includegraphics[width=0.45\textwidth]{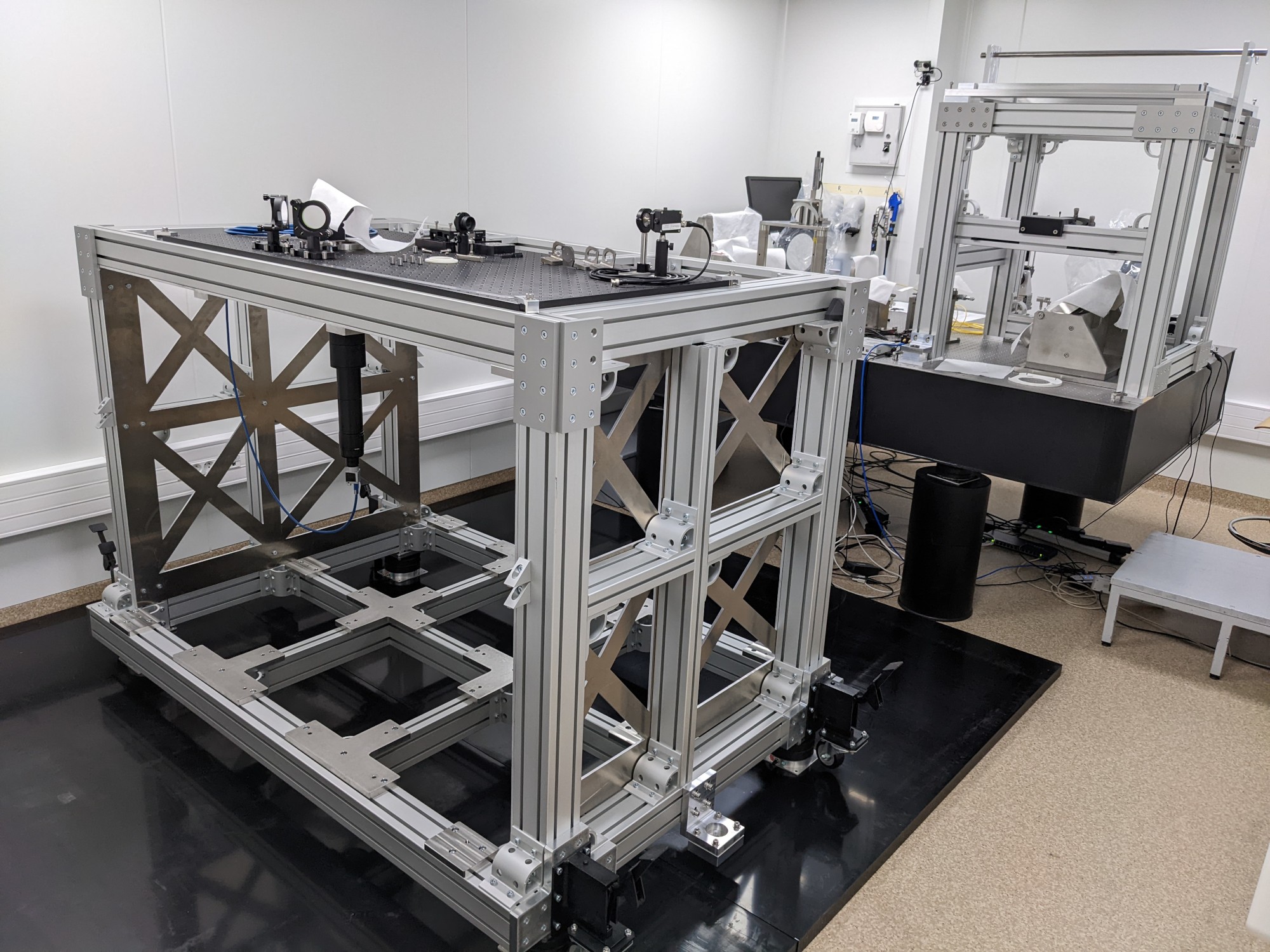}
    \includegraphics[width=0.45\textwidth]{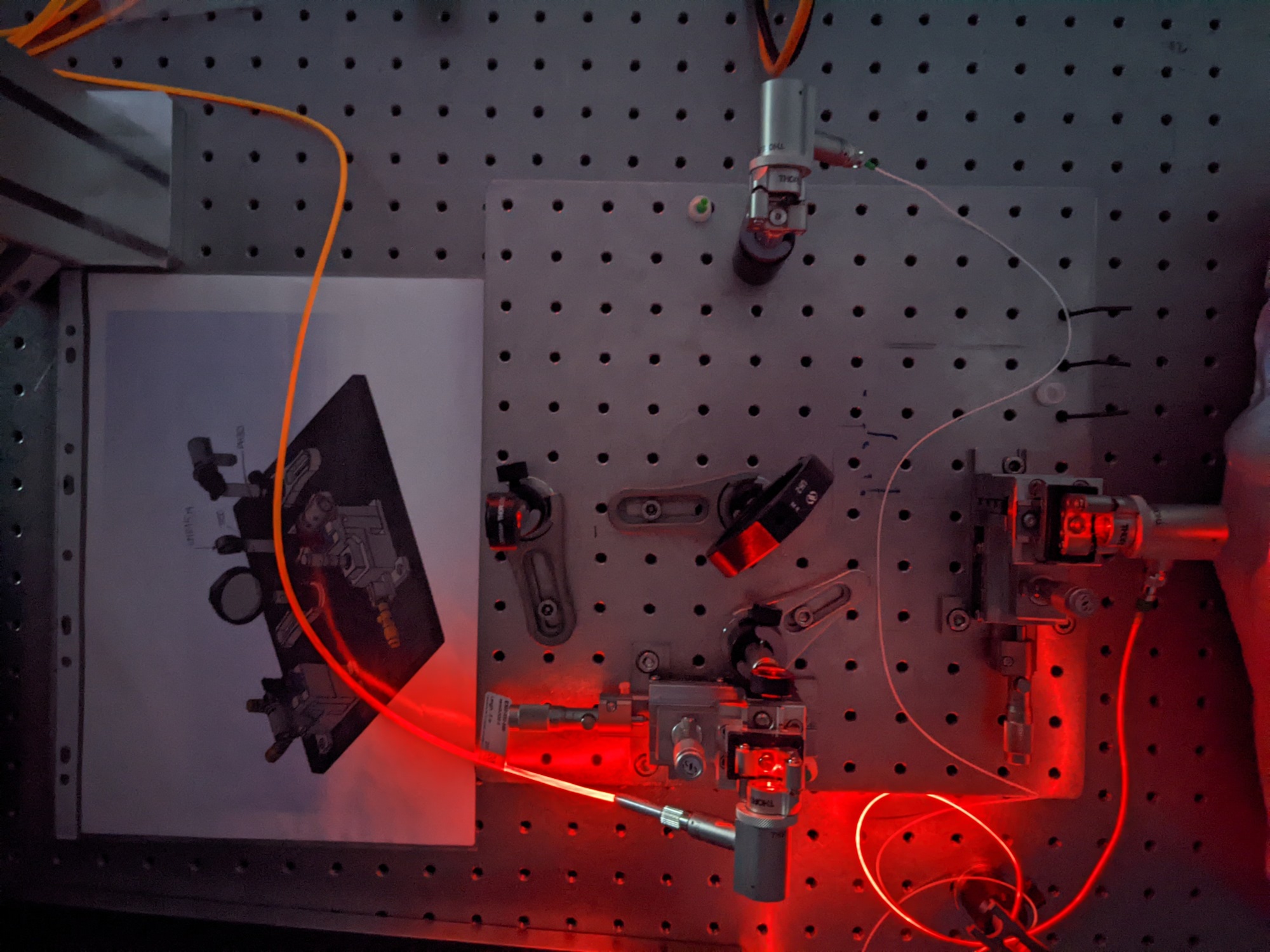}
    \caption{{\bf Left:} GPAO testbench as built in the clean room.  {\bf Right:} Picture of the working source (note the LGS part is being mounted right now and is therefore not visible here). }
    \label{fig:asbuilt}
\end{figure}

The test bench is now fully installed (Figure \ref{fig:asbuilt}, left) and is being characterized to verify its specifications (optical characteristics, stability).
The NGS source is installed and working (Figure \ref{fig:asbuilt}, right). The installation of the LGS source is ongoing.

\begin{figure}[htbp]
    \centering
    \includegraphics[width=0.32\textwidth]{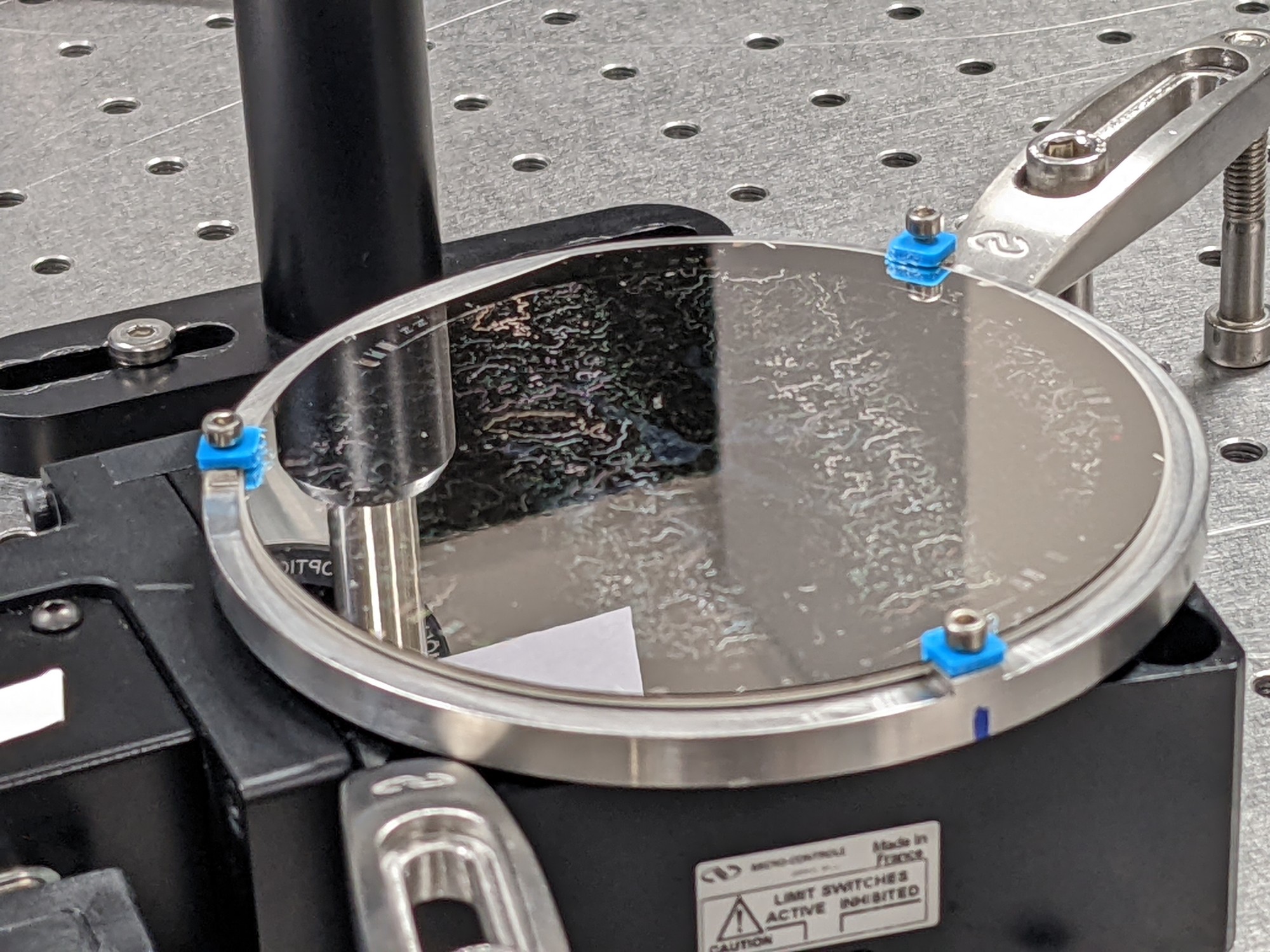}
    \includegraphics[width=0.32\textwidth]{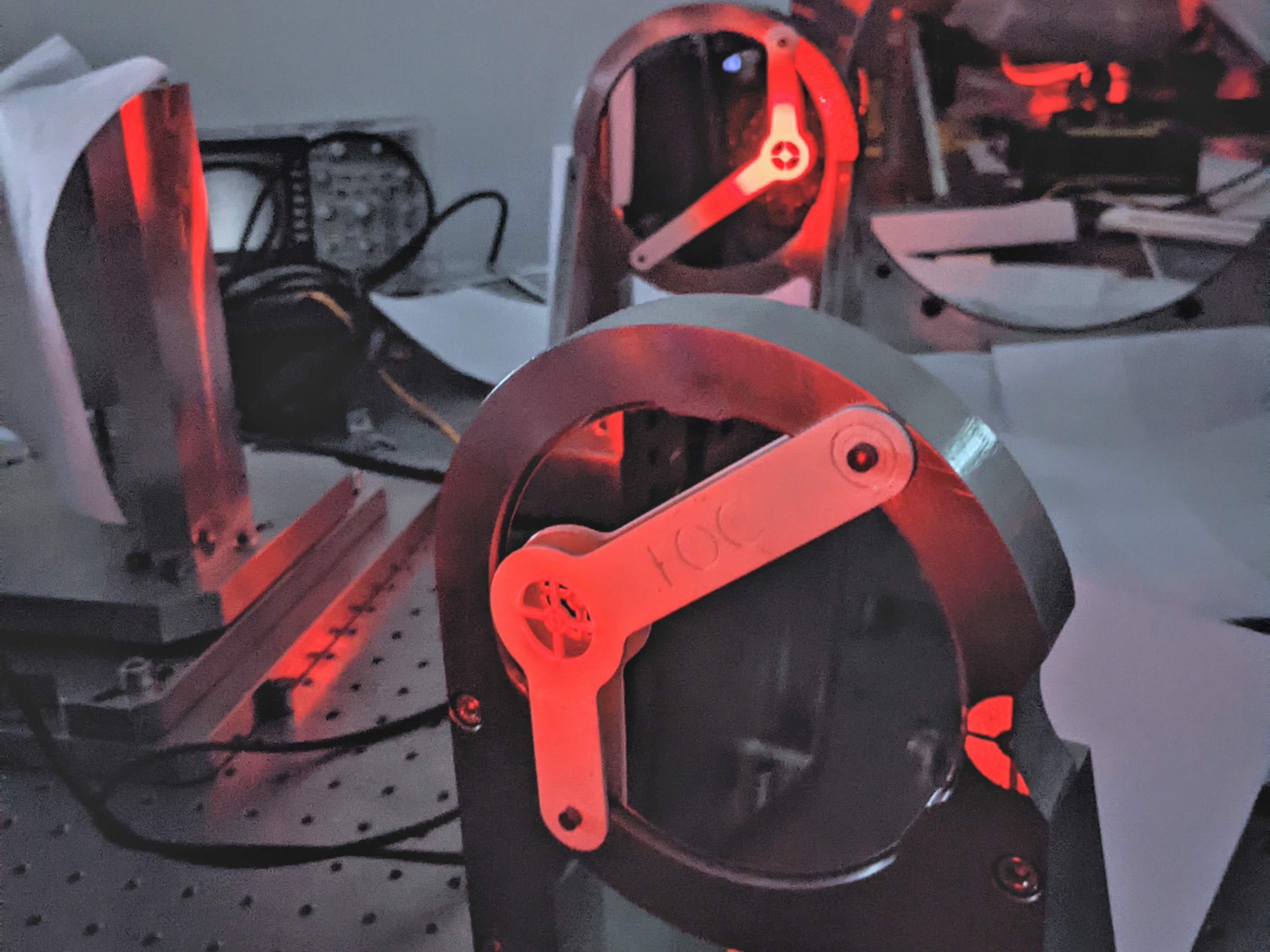}
    \includegraphics[width=0.32\textwidth]{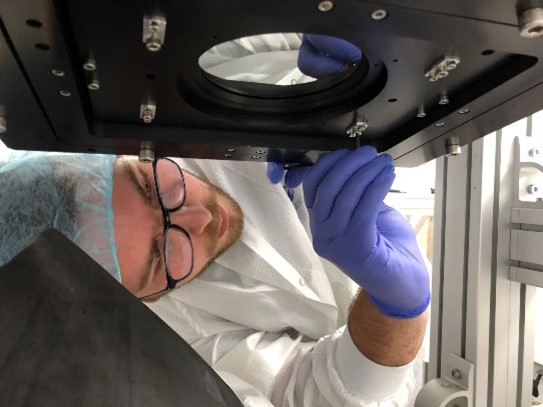}
    \caption{{\bf From left to right:} SILIOS phase plate installed on its rotating mount. 
    3D-printed masks used for the alignment of the testbench. Installation of the Corrective Optics in Nice with its "dummy" flat mirror that replaces the Deformable Mirror for optical alignments.}
    \label{fig:phasemask}
\end{figure}

The bench turbulence simulator (phase plate) is also operational: a phase screen in reflection (Figure \ref{fig:phasemask}, left) from SILOS technologies can be rotated in order to introduce an achromatic turbulent phase in the beam, leading to speckle patterns. The phase map inscribed onto the phase plate was designed so as to produce the equivalent of a 0.8" seeing for a 8m-like telescope.
This phase plate can be replaced by a flat mirror for alignment and characterization of the bench.

For the alignment of the test bench, we made use of the CAD design (Figure \ref{fig:schematics}), where we have all the light beams materialized on top of our opto-mechanics design, to materialize them onto 3D-printed masks that we fixed in front of each optical element (Figure \ref{fig:phasemask}, middle). Aligning the bench ended up in placing the beams on their respective 3D-printed targets, ensuring a sub-mm precision prior to any fine-tuning.
This procedure happened to ease greatly the alignment of such a complex system, and reduced the usual overhead between mounting and operation of the bench.

A flat "dummy" mirror has been delivered by IPAG, Grenoble to Lagrange, Nice and mounted in the final DM structure to finalize the alignment of the bench and to perform temperature behavior tests (Figure \ref{fig:phasemask}, right). More details on these tests can be found in Nowacki et al. 2022\cite{Nowacki2022}. This also allowed us to finalize the bench alignment up to the WFS focus and to verify the bench specifications.

\section{Some characteristics of the GPAO testbench} \label{sec:characteristics} 

The main GPAO testbench specification is to have an output beam identical to the one of the Unit Telescope beam, i.e. a light beam that has a 10\,cm diameter onto the deformable mirror (convergent beam), and a F-number (or $\frac{F}{D}$ number) of 46.7, $F$ being the focal length of the telescope, and $D$ being its diameter. The beam size at the place of the DM is checked visually using masks that can be put on top of the mirror.

The $\frac{F}{D}$ number can be easily verified using Point Spread Functions (PSF) images from the testbench. We calculate $\frac{F}{D} = \frac{{\rm FWHM} * {\rm Pix}}{\lambda}$ where FWHM is the Full Width at Half Maximum of the PSF, Pix the pixel scale of the camera, and $\lambda$ the wavelength of the source. We measured the FWHM of the PSF  and we verified this way that the main bench specification is met by $0.5\%$: the measured $\frac{F}{D}$ number is 46.95 ± 0.03.



\begin{figure}[htbp]
    \centering
    \includegraphics[width=0.31\textwidth]{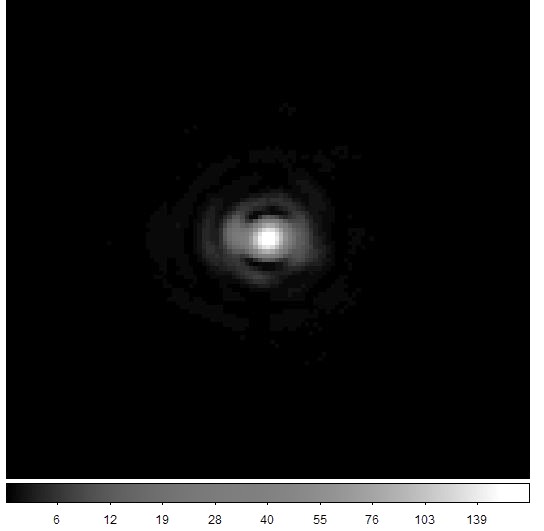}
    \includegraphics[width=0.35\textwidth]{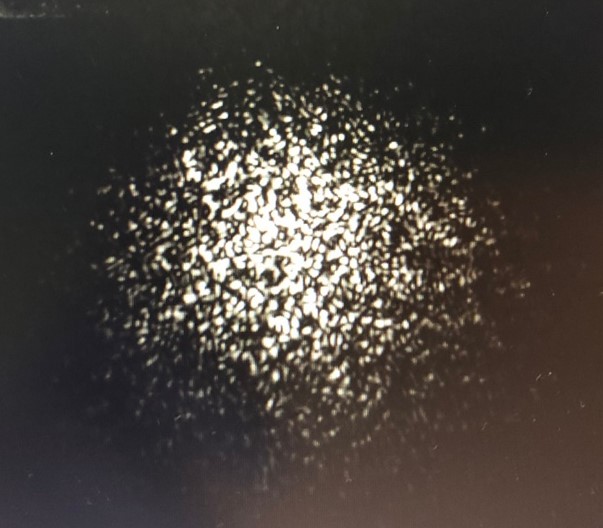}
    \includegraphics[width=0.305\textwidth]{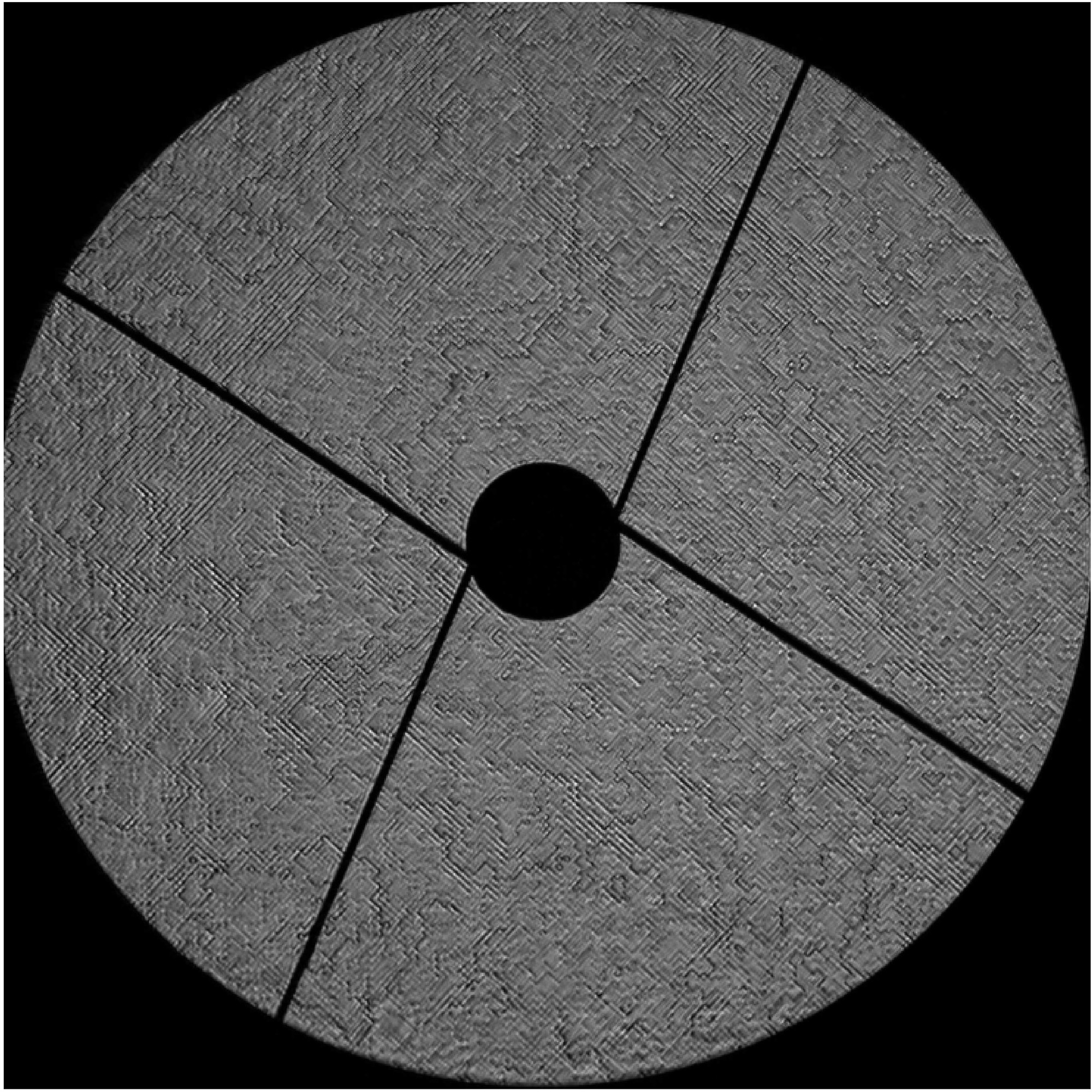}
    \caption{{\bf Left:} GPAO testbench PSF core (hyperbolic arc-sinus scale), measured with the visible bench camera. {\bf Middle:} visible PSF with the phase mask in place (screen shot). {\bf Right:} Pupil vizualisation, showing the UT mask (in black) in front of the phase mask (giving an uniform grainy pattern).}
    \label{fig:psfpup}
\end{figure}

One of the preliminary bench PSF is shown in Figure \ref{fig:psfpup} (left), showing a healthy image quality although showing here a small astigmatism. This astigmatism evolving with time, we decided to act on two offending glued flat mirrors, that have been unglued and instead clamped in place in their mounts to correct for this effect.

The work is now focused on characterizing the phase mask and verify it meets the specifications provided to SILIOS technologies. After a visual microscope inspection, and phase measurements into a Zygo interferometer, it has been installed onto the bench. One "turbulent" PSF is shown  in Figure \ref{fig:psfpup} (middle) for illustration.

A pupil-visualization lens can be put in place onto the breadboard supporting the bench cameras. The pupil obtained this way can be seen in in Figure \ref{fig:psfpup} (right), when both the pupil mask and the phase plate are in place in the bench. The pupil mask reproducing the central obstruction as well as the spider of an UT can be seen as a black mask, and the phase mask itself exhibits its pixellated grainy surface, result of the manufacturing process of SILIOS technology.

\begin{figure}[htbp]
    \centering
    \includegraphics[width=0.45\textwidth]{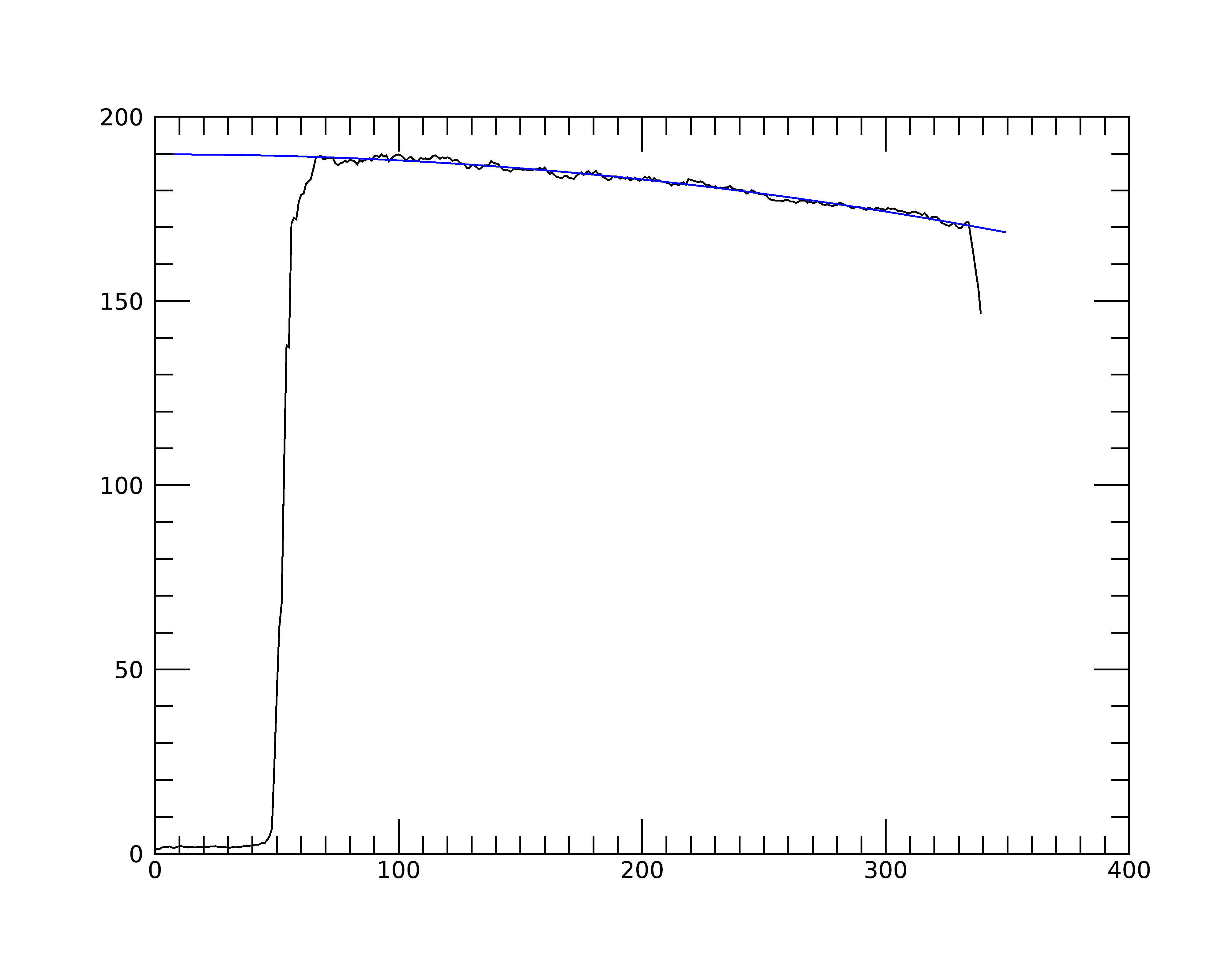}
    \includegraphics[width=0.45\textwidth]{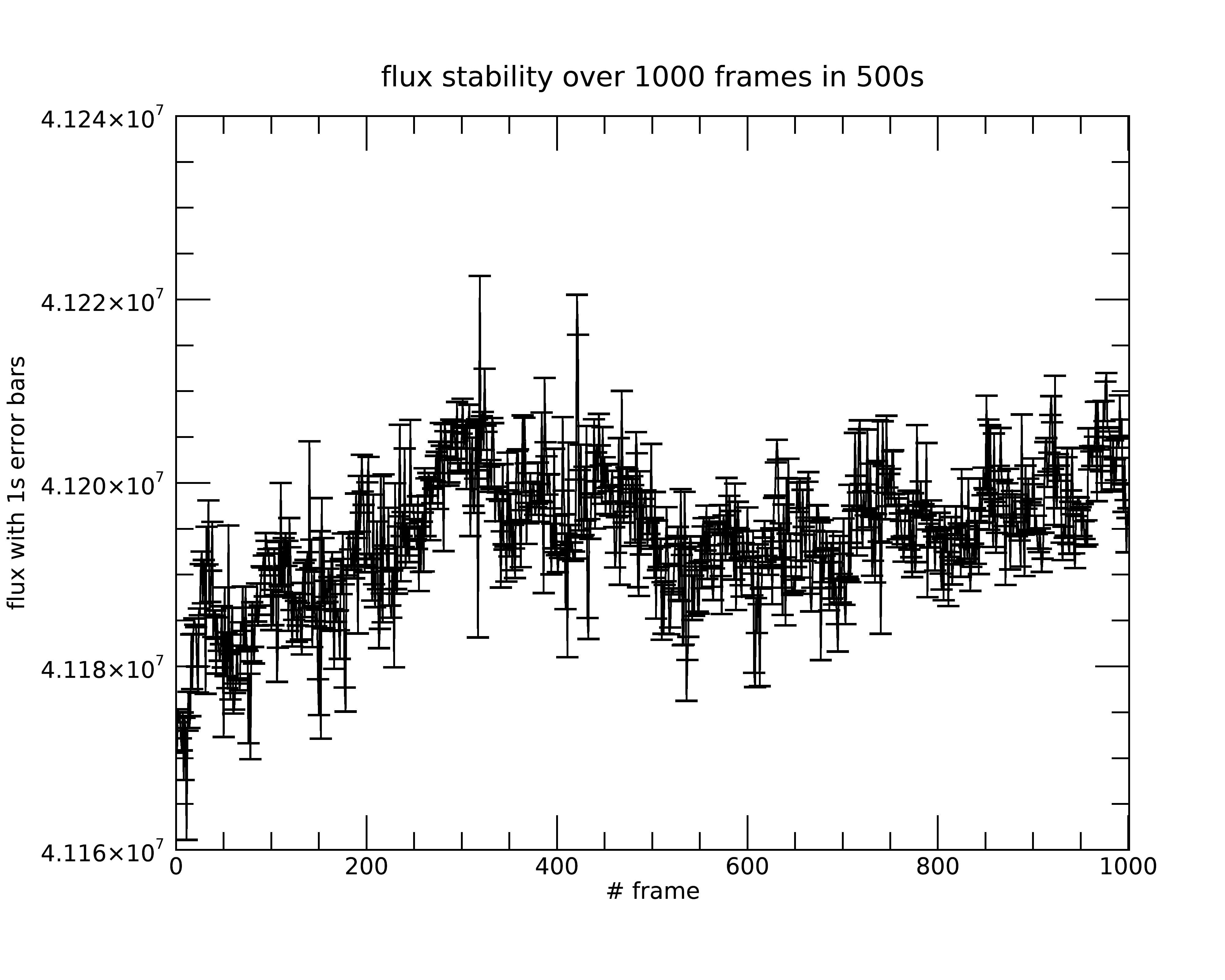}
    \caption{{\bf Left:} Pupil illumination homogeneity, showing an azimuthally-integrated pupil flux (in black, arbitrary units) compared to a polynomial fit (blue curve) as a function of detector pixel number. {\bf Right:} Flux averaged over the full pupil as a function of time.}
    \label{fig:illumstab}
\end{figure}

We present finally two additional characteristics of the testbench tailored to prepare the venue of all the GPAO subsystems in it: we measured the illumination homogeneity across the pupil of 0.8 (borders) / 1 (center), meeting its specifications (Figure \ref{fig:illumstab}, left), and finally we measured the flux stability in visible light of the bench over roughly 8\,mn, showing sub-percent stability over that period of time.

\section{GPAO test plan} \label{sec:testplan}

Once all the bench specifications have been met, we will start the test plan of GPAO. It will consist in bringing all the AO subsystems (deformable mirror, wavefront sensor, real time computer) onto the testbench at the end of 2022, have them work together through the Real-Time Computer, and perform a series of open loop and closed loop runs to verify the AO perfromances are similar to the expected ones. But this will be presented in a next episode!

\acknowledgments 

GRAVITY+ is being built by a consortium composed of German (MPE, MPIA and University of Cologne), French (CNRS-INSU: LESIA, Paris, IPAG, Grenoble, Lagrange, Nice, and CRAL, Lyon), British (University of Southampton), Belgium (KU Leuven), Portugese (CAUP) institutes. It was defined, funded and built in close collaboration with ESO. We also thank Observatoire de la Côte d'Azur and Université Côte D'Azur that provided resources to manufacture the testbench. F. Millour acknowledges funding from the \emph{Agence Nationale de la Recherche} (ANR) in the EXOVLTI project ANR-21-CE31-0017-03. French partners thank the specific action ASHRA and the national programs PNP, PNPS and PNCG for their support in the GRAVITY+ project.

\bibliography{report} 
\bibliographystyle{spiebib} 

\newpage




\end{document}